\documentstyle[aaspp]{article}

\newcommand{\beq}{\begin{equation}} \newcommand {\eeq}{\end{equation}}
\newcommand{\beqa}{\begin{eqnarray}}
\newcommand{\eeqa}{\end{eqnarray}} 
\newcommand{\half}{H\alpha}

\newcommand{\teff}{T_{eff}}
\newcommand{\mvo}{M_{Vo}}
\newcommand{\ebv}{E(B-V)}
\newcommand{\bvo}{(B-V)_o}

\newcommand{\averebv}{\overline{E(B-V)}}

\begin{document}
\title {ARE GALAXIES OPTICALLY THIN \\TO THEIR OWN LYMAN CONTINUUM
RADIATION?\\ I. M33}
\author{Kanan Patel and Christine D. Wilson}
\affil{Department of Physics and
Astronomy,\\McMaster University,\\Hamilton, Ontario L8S 4M1 Canada}

\begin{abstract}

  Previously published $\half$ data and UBV photometry of blue stars in the
inner kpc of M33 are used to study the distribution of OB stars and HII
regions in the galaxy and to determine whether individual regions of the
galaxy are separately and/or collectively in a state of ionization balance.
Based on the surface brightness of the $\half$ emission, we identify three
distinct ionized gas environments (bright, halo and field). We find that
$\sim$50\% of the OB stars are located in the field, so that 1/2 of the
lifetime of OB stars must be spent outside recognizable HII regions.  If OB
stars escape from bright HII regions by destroying their parent molecular
clouds, this result would imply that molecular cloud lifetimes after forming
OB stars could be as low as $\sim$5$\times$10$^6$ yrs or 1/2 the typical
lifetime of OB stars.  We show that a possible origin for the large field OB
population is that they were born in and subsequently percolated out of the
$\sim$10$^3$ giant molecular clouds with masses $\ge$10$^3$M$_\odot$
predicted to exist within the inner kpc of the galaxy.  Using ionization
models, we predict $\half$ fluxes in the bright, halo and field regions and
compare them to those observed to find that the regions, separately as well
as collectively, are not in ionization balance:  predicted fluxes are a
factor of $\sim$3-7 greater than observed.  The heaviest loss of ionizing
photons appears to be taking place in the field.  Observed and predicted
$\half$ luminosities in the field are in best agreement when Case A
recombination is assumed. Therefore, our findings suggest that star
formation rates obtained from $\half$ luminosities must underestimate the
true star formation rate in these regions of M33.  We have performed a
similar analysis of an individual, isolated region with bright and halo
$\half$ emission to find that comparable results apply and that the region,
as a whole, is also not in ionization balance.

\end{abstract}

\clearpage


\section{INTRODUCTION}

Although massive OB stars have long been recognized as the source of the
ionizing radiation that defines  optically luminous HII regions, only
recently has it become clear that OB associations and HII regions do not
always coincide spatially. The Local Group galaxy M33 contains many examples
of extragalactic OB associations that do not contain bright HII regions
(Wilson 1990) while the edge-on spiral NGC 891 contains faint, filamentary
$H_\alpha$ structures high above the galactic disk with no obvious ionizing
sources (Rand et al.  1990). In addition, the luminous young stars in M33
are not restricted to the spiral arms, traditionally believed to be the
primary site of star formation (Madore 1978). Studies of the OB associations
in the Magellanic Clouds show that there is an entire population of massive
young stars that are located far enough away from neighbouring OB
associations that they must have been born in the \lq\lq field" and not in
the associations (Massey et al. 1994).  These observations suggest that
massive stars must spend some portion of their lives far from dense gas
clouds and hence outside bright HII regions.  The presence of significant
numbers of field O stars has important implications for a number of key
issues in star formation and the interstellar medium, including the
evolution of molecular clouds, the ionization of diffuse interstellar gas
and the calculation of extragalactic star formation rates (SFRs).

The length of time that an O star spends in the HII region phase, and thus
at least partially embedded in its parent gas cloud, is important for
estimating molecular cloud lifetimes and, therefore,  understanding the
evolution of the molecular interstellar medium in galaxies.  Estimates of
molecular cloud lifetimes require an assessment of the destructive
influence of massive stars and thus the time such stars spend in their natal
environment is very important. The short lifetimes of $\sim$10$^7$ yrs
obtained by some authors (cf. Blitz \& Shu 1980) assume this embedding time
is a large fraction of the total lifetime of the star so that the first O
stars that form will disrupt their parent cloud.  Longer lifetime estimates
($\geq$10$^{8}$ yrs, Solomon \& Sanders 1980) are based on clouds being
formed in a stable environment, the spiral interarm region of the Galaxy,
where the overall SFR is low.  Molecular clouds undergoing periods of
intense star formation followed by quiescent star formation would result in
long molecular cloud lifetimes so that each cloud could produce several OB
associations of different ages (Elmegreen 1991).  The fraction of
self-gravitating clouds currently experiencing star formation ($\sim$0.5)
can be combined with the duration of the process ($\sim$20$\times$10$^6$yrs
in OB associations) to obtain molecular cloud lifetimes of
$\sim$4$\times$10$^7$yrs (Elmegreen 1991).  If OB stars destroy their natal
environments, then the fraction of O stars seen in the field determines the
time required to destroy the cloud.  However, if these stars are seen in the
field because they simply moved out of the clouds, then cloud lifetimes could
be considerably longer.

The presence of OB stars in the field may be important to understanding the
origin of the diffuse ionized gas (DIG)  observed in our own Galaxy as well as
extragalactic systems.  This gas is characterized by its low density
(electron density $n_e\sim$0.2 cm$^{-3}$), cool temperatures
($<$10,000 K) and high [SII]/$\half$ line intensities (Walterbos
1991).  Observations of edge-on galaxies indicate scale heights of the
DIG  can be of the order of a few kpc (Walterbos 1991).
Diffuse ionized gas in external galaxies is believed to account for a
substantial fraction of the total $\half$ emission from a galaxy (20-30\% in
the Large Magellanic Cloud, Kennicutt \& Hodge 1986; 20-40\% in M31,
Walterbos \& Braun 1994).  Currently the most likely sources of ionizing
photons for the DIG  are OB stars.  One  problem with
this scenario is that there is no clear understanding of how ionizing
photons can traverse the large distances above the plane.  If significant
numbers of OB stars are found in field regions (where $\half$ emission is
low and optical depths are likely to be smaller than in bright HII regions),
a potentially large reservoir of ionizing photons would be available to
power the DIG.

 Star formation rates that are calculated from observed $\half$ fluxes rely
on the assumption that the region in question is {\it ionization bounded}
(Kennicutt 1983) i.e. all Lyman continuum photons produced by massive stars
are absorbed by the gas in the region (or galaxy). (The alternative scenario
is for a region (or galaxy) to be {\it density bounded}, i.e. the Lyman
photons are more than sufficient to ionize all the (dense) gas.)  The
observed $\half$ luminosity is then directly proportional to the Lyman
continuum
luminosity which can be converted into a SFR for massive stars via an
assumed  initial mass function (IMF) and theoretical
models for the Lyman
continuum luminosity as a function of stellar mass.  The total SFR
of a region can be  obtained by extending the  IMF over the full range of
stellar masses.  In these calculations, the assumption that the region is
ionization bounded is crucial: since  the SFR is directly proportional to the
Lyman continuum, and therefore $\half$, luminosity, a region out of which
continuum photons are escaping will have a smaller estimated SFR than one
that is in ionization balance.  The assumption of ionization boundedness
may not be valid if large portions of the OB population lie outside bright
HII regions. Considering the wide spread use of this technique for obtaining
SFRs in extragalactic systems, it is vital to verify this underlying
assumption of SFR measurements from $\half$ luminosities.

 Ionization balance calculations in extragalactic OB associations and HII
regions have concentrated on single HII regions in the Large and Small
Magellanic Clouds (LMC and SMC).  Massey et al. (1989a) find the LMC OB
associations NGC 2122 and LH 118  to be density  rather than
ionization bounded, since the predicted fluxes from stars in the association
are
a factor of two greater than what is observed.  A similar study of NGC 346
in the SMC (Massey et al. 1989b) found the opposite to be true:  the models
predict less flux (up to a factor of $\sim$2) than what is observed and so the
region is most likely ionization bounded.  Finally,  Parker et al. (1992)
found that the LMC OB associations LH 9 and LH 10 were at the limit of being
ionization bounded (observed fluxes were at most a factor of 1.5 greater
than those  predicted). To date there have been no such studies of systems
more distant than the Magellanic Clouds and no attempt to measure the
ionization boundedness of a large area of a galaxy.

 At a distance of 0.79 Mpc (van den Bergh  1991), the Local Group galaxy M33
is an ideal system in which to conduct such an investigation.  This Sc type
galaxy is undergoing vigorous high mass star formation.
The inner kpc region of the galaxy has recently received much attention:
molecular gas maps in CO have identified a population of  molecular
clouds (Wilson \& Scoville 1990), optical imaging at $\half$ has provided
quantitative photometry of
the HII regions (Wilson \& Scoville 1991), and a UBV photometric survey of
the region has identified luminous blue stars and OB associations (Wilson
1991). Diffuse ionized gas within the galaxy has been investigated by
Hester \& Kulkarni  (1990).

In this paper, we use $\half$ and UBV photometry of the inner kpc region of
M33 to study the distribution of the HII regions and luminous blue stars
(hereafter OB stars) so as to calculate the fraction of an OB star's lifetime
that is spent outside an HII region and to address the issues of molecular
cloud lifetimes, ionization balance, and the implications for the
calculation of star formation rates and powering the diffuse ionized gas.
The analysis of the $\half$ data and UBV photometry and the theoretical
ionization models used are outlined in section \S 2. A discussion of the OB
star distribution is given in \S 3 and the results of the ionization balance
calculations appear in
\S 4.  The results of our investigation are summarized in \S 5.

\section[Selection and Analysis  of Data]
{SELECTION AND ANALYSIS  OF DATA}

\subsection{Photometric Data}

We use the photometric UBV data from Wilson (1991), who used it to study the
OB associations in the nucleus of M33.  The observations were made at the
Palomar 60-inch and Canada-France-Hawaii telescope (CFHT), using the blue
sensitive Tektronix chip, CCD 6 (scale 0.235$^{\prime\prime}$ pixel$^{-1}$)
at Palomar and the RCA2 double density CCD chip (scale
0.205$^{\prime\prime}$ pixel$^{-1}$) at the CFHT.  The seeing was
1.2$^{\prime\prime}$-1.9$^{\prime\prime}$ at Palomar and
$\sim$0.8$^{\prime\prime}$ at the CFHT.  Additional details of the
observations can be found in Wilson (1991).  Systematic uncertainties in the
photometric zero points were estimated to be $\pm$0.05 mag in B, $\pm$ 0.04
mag in V and $\pm$0.15 mag in U.  The data were estimated to be incomplete
by 30\% for 19.5$<$V$<$20, 55\% for 20$<$V$<$20.5 and 65\% for
20.5$<$V$<$21.  Subsequent comparison of the data with independent
photometry of M33 obtained by P.  Massey (private communication) showed that
there is a systematic offset (Massey-Wilson) of 0.14 mag in (U-B), while the
B and V magnitudes agreed to better than 0.04 mag. Hence we have applied a
+0.14 mag correction to the (U-B) colours, i.e. the colours are shifted to
the red relative to the values published in Wilson (1991).

 This data set is particularly suitable for our use as it provides the most
complete (to 21 mag in V) large-area census of the blue star population in
the inner regions of the galaxy. The average reddening (foreground and
internal) for the surveyed region is $\averebv$=0.3$\pm$0.1 mag (Wilson
1991). We selected only those stars with \bv$\leq$0.4 mag.  This criterion,
which is based on the estimated average reddening for the galaxy, the
intrinsic colour of O stars on the the zero age main sequence (ZAMS),
$\bv$=-0.3 mag (Flower 1977), and the observed width of the main sequence,
assures us that most potential OB stars will be selected.

 \subsection{$\half$ Emission}

We used a red-continuum subtracted H$_\alpha $ image of M33 from Wilson \&
Scoville (1991) to study the $\half$ emission in the galaxy. The image was
obtained on the Palomar 60-inch telescope (rebinned chip scale
1.3$^{\prime\prime} $pixel$^{-1}$).  Visual inspection of the continuum
subtracted $\half$ image (henceforth the $\half$ image) strongly suggested
that three types of HII regions, bright, halo and field, are present in
M33.  The different regions were defined by their H$_\alpha$ surface
brightness, {\it I}, such that {\it I} $\geq$1.37$\times$10$^{-15}$ erg
s$^{-1}$cm$^{-2}$ arcsec$^{-2}$ for the bright regions,
3.61$\times$10$^{-16}\leq${\it I}$<$1.37$\times$10$^{-15}$ erg s$^{-1}$
cm$^{-2}$ arcsec$^{-2}$ for the halo regions and {\it I}$<$3.61
$\times$10$^{-16}$ erg s$^{-1}$ cm$^{-2}$ arcsec$^{-2}$ for the field
regions (see Figure 1a).  The divisions into the three regions are
motivated purely by visual inspection of the total image; the adopted
dividing lines in surface brightness between each type of region are not
based on any intrinsic physical quantity.  The surface brightnesses, which
have already been corrected for extinction in $\half$
(A$_{\half}$=2.59$\times\ebv$=0.78 mag (Schild 1977)) were measured above
an observed average sky surface brightness over the galaxy of
I$_{bg}$=5.43$\pm$0.10$\times$10$^{-16}$ erg s$^{-1}$ arcsec$^{-2}$.  The
latter was obtained by averaging the observed surface brightness of four
broad regions of the galaxy that appeared free of any $\half$ emission.
Because our $\half$ image of M33 is restricted only to the inner kpc of the
galaxy and does not cover any areas far from the galaxy, we cannot be
certain that the background emission is in fact
\lq\lq sky", and not a very smooth emission component in M33. We will be in a
better position to determine the true source of this background emission in
our upcoming study of NGC 6822, for which we have $\half$ images that extend
past the optical limits of the galaxy.  Finally, we note that these $\half$
surface brightnesses as well as all measures of the $\half$ luminosity have
been corrected for contamination by the [NII] 6583$\AA$ line that was also
observed through the filter used to obtain the $\half$ image.  To correct
for this contamination, we used the data of Vilchez et al. (1988) to
determine the average [NII]/$\half$ line ratio of 21\% $\pm$1\% in three HII
regions that lie within the surveyed region of M33. Scaling down all
counts in the $\half$ image by a factor of 0.79 then corrects for this
contamination in the data.

The uncertainties associated with the observed $\half$ luminosity include
those in the calibration of the $\half$ emission and the extinction at
$\half$.  Wilson and Scoville (1991) estimate the calibration to be accurate
to $\pm$5\%. The uncertainty in A$_{H\alpha}$ was determined to be
A$_{H\alpha}$=0.78$\pm$0.26 mag from the assumed mean reddening for the
galaxy $\averebv$=0.3$\pm$0.1 mag and the uncertainty in the conversion
factor between A$_{\half}$ and $\ebv$, which we estimated at 0.07 mag or
$\sim$10\% of A$_{\half}$.  We note that a radio continuum survey of the HII
regions in M33 found an average A$_{\half}$ of 0.91 mag (Viallefond
\& Goss 1986) which is comparable to our estimate. The total uncertainty in
the observed $\half$ luminosities is thus $\sim$30\%.
\nolinebreak

\subsection{Predicted $\half$ Luminosity from OB stars}

In order to compare the observed $\half$ fluxes with those expected from the
associated stellar population, we estimate the Lyman continuum flux, and
hence the $\half$ luminosity, for stars with (B-V)$\leq$0.4 by applying a
number of stellar ionization models: blackbody, Auer \& Mihalas (1972)
(hereafter A\&M), Kurucz (1979) and Panagia (1973).  Both A\&M and Kurucz
calculate the expected number of Lyman continuum photons as a function of
the effective temperature, $\teff$, for various values of the effective
gravity, $g$.  The A\&M results are based on nonblanketed NLTE model
atmospheres composed of hydrogen and helium only.  The Kurucz models, which
include the effects of line blanketing, have LTE atmospheres of solar
composition. The Panagia models are based on a combination of the NLTE A\&M
models and the LTE models due to Bradley and Morton (1969), Morton (1969)
and Van Citters \& Morton (1970), and appropriate values of $g$ are used to
calculate the flux of Lyman continuum photons as a function of $\teff$ for
stars of luminosity class I, III, V as well as the ZAMS.

Effective temperatures ($\teff$) and bolometric corrections (BC) for stars
with M$_V$$\leq$21 and $\bv$$\leq$0.4 were obtained using the calibration
equations given in Parker \& Garmany (1993).  The six equations (4a to 4f in
Parker \& Garmany 1993) give $\teff$ in terms of the reddening free index
Q$\equiv$(U-B)-0.72(B-V) and the dereddened colour, $\bvo$, for various
luminosity classes. These equations which are, in large part, taken from
Massey et al. (1989a), are based on the data of Flower (1977), FitzGerald
(1970) and Humphreys \& McElroy (1984).  The BC of the star was determined
from $\teff$ using one of equations 5a to 5d in Parker \& Garmany (1993).
These calibrations are from Chelbowski \& Garmany (1991) and Flower (1977).
In applying these calibration equations, we followed the basic procedure
outlined in Parker  \& Garmany,  with a few modifications.  We first selected
all stars with M$_V\!\leq$-6 as potential supergiants. If the star had a
(U-B) colour, we used equation (4f) to calculate $\teff$. For
-1.1$<$Q$\leq$-0.9 or Q$>$0.0, we used equation (4a) (which is valid for
main sequence stars) and subtracted 4000 K from the calculated $\teff$ (as
per Parker \& Garmany 1993; see their section 3.1 for a discussion of this
point). For stars with Q$<$-1.1 or -0.3$<$Q$\leq$0.0, we used (i) equation
(4b) -4000 K if $\bvo$$<$0.0.  (ii) equation (4c) if 0.0$\leq\bvo$$<$0.2 or
(iii) equation (4d) if 0.2$\leq\bvo$$<$0.5.  Effective temperatures for all
remaining, non-supergiant stars were calculated using one of equations (4a)
to (4e) depending on the value of Q (if available) and $\bvo$.

 For the blackbody, A\&M and Kurucz models, we applied Table XIV of Massey
et al. (1989), which neatly summarizes the expected Lyman flux at the surface
of the star for $\teff$ between 20,000 K and 50,000 K.  For the
Kurucz models, we have used log($g$)=3.5 for stars with $\teff\leq$
35,500 K, and log($g$)=5.0 for stars with $\teff>$35,500 K
while for the A\&M models we take log($g$)=4.0.  The total flux in Lyman
continuum photons is then found by integrating over the surface area of the
star.  For each star for which $M_{bol}$ and $\teff$ were determined (as
outlined above) and 18,000 K$\leq\teff\leq$60,000 K, we
calculate the stellar radius, R$_*$, using the luminosity-effective
temperature relation, L=4$\pi$R$_*^2\sigma_{B}\teff ^4$.

  To apply the Panagia models, we need to know the luminosity class of the
star as well as its $\teff$.  Panagia gives visual magnitudes (M$_v^P$) as a
function of effective temperature (16,100 K$\leq$$\teff ^P \!\leq$50,000 K)
for ZAMS, supergiant, class III and main sequence stars.  We estimate the
luminosity class of the star by comparing the observed visual magnitude
corrected for foreground and internal extinction, $\mvo$, with M$_v^P$.
Only stars with 15,100 K$\leq$$\teff$$\leq$$T_{eff, U}\!\equiv$ 60,000 K
are considered.  We take all stars with $\mvo\!\leq -$6.0 to be supergiants
(class I).  For the remainder of the stars, we compare the observed $\mvo$
to the values of the three M$_v^P$'s (for class III, V and ZAMS)
corresponding to the star's $\teff$.  The M$_v^P$ that is closest to the
observed $\mvo$ determines the luminosity class. In turn, the luminosity
class and $\teff$ gives the total flux of Lyman continuum photons from the
entire star.

  To convert Lyman continuum photons to $\half$, we refer to Osterbrock
(1989) to find N(LyC)/N($\half$)=2.22 assuming Case B recombination
(appropriate for optically thick gas at 10,000 K and densities of
10$^2$-10$^4$ cm$^{-3}$).  Although Case B recombination is the most
commonly used approximation in the study of HII regions, we also consider
Case A recombination for which N(LyC)/N($\half$)=5.36 (Brockelhurst 1971),
for stars in the \lq\lq field" where the optical depth of the surrounding gas
may be smaller.  Clearly, Lyman continuum photons in optically thin gas
produce fewer $\half$ photons than those in optically thick gas.

\section[Ionized Gas and OB Star Distribution] {IONIZED GAS AND OB STAR
DISTRIBUTION}

O and B stars in galaxies are commonly assumed to be located in bright HII
regions. To study this assumption quantitatively, we identified the ionized
gas environment of a star as either bright, halo, or field according to the
average $\half$ surface brightness inside a circle of radius of 2 pixels
($\sim$9.8 pc) centered on the star.  Table 1 summarizes the
total numbers of stars as a function of magnitude found in the three
different ionized gas environments.

 Inspection of these tables reveals a number of surprising results.  First
and foremost, we find that {\it all O stars are not in bright HII regions}:
there are a significant number of O stars in the field (see Figure 1b).
 From our analysis, roughly 50\% of all blue ($\bv$$\leq$0.4) stars brighter
than V=21 are located in the field. In fact, the number of field stars
outnumbers those in bright HII regions by almost a factor of three.  This
result is reflected also in the brightest stars (V$\leq$18), where almost a
third of the total are field stars.  The halo component of the $\half$
emission also contains a significant percentage ($\sim$30\%) of the stellar
population.

 Additionally, the bright regions are predominantly populated by faint
stars:  50\% of all stars in the bright regions have 21$<$V$\leq$20
compared to 6\% with V$\leq$18.  Although it appears that fainter stars
form a larger fraction of the population in the field than in the bright HII
regions ($\sim$64\% in the field compared to $\sim$50\% in the bright
regions), this  may be due to a higher incompleteness level in the
brighter regions where the crowding is larger (see Table 1).

 From these results, we can estimate the fraction of the main sequence
lifetime of an O star that is spent within a recognizable HII region
($f_{HII})$.  We consider such an object to be composed of either a bright or
halo   emission region so as to obtain the most conservative estimate of
this important quantity.  With this assumption, we obtain $f_{HII}\equiv
(N_B+N_H)/N_{TOT}$=0.5 where $N_{TOT}$ is the total number of stars in the
sample and $N_B$ and $N_H$ are the numbers of stars in, respectively, the
bright and halo  regions of the galaxy. Thus roughly 50\% of the main
sequence lifetime of an O star is spent {\it outside } of a recognizable HII
region. If OB stars escape from bright HII regions by destroying their
parent molecular clouds, this result would imply that  molecular cloud
lifetimes after forming OB stars could be as low as $\sim$1-4$\times$10$^6$
yrs (or 1/2 the typical MS lifetimes of 2.6-8.1$\times$10$^6$yrs for 120-20
M$_\odot$ stars assuming Z=.020, Schaller et al. 1992).  Since 50\% of the
molecular clouds in M33 contain optically visible HII regions (Wilson \&
Scoville 1991), under this scenario the lifetimes of the molecular clouds
would be $\le$10$^7$ yrs. This  result, however,  is inconsistent with the
evidence that some OB associations in the galaxy have undergone at least two
episodes of star formation separated by 10$^7$ yrs (Regan \& Wilson 1993).

An alternative scenario is that the field O stars have percolated out of
existing bright HII regions.  In Table 2 we summarize a few
key properties of the O star distribution and ionized gas environments. We
see that although the field region occupies an area about 10 times that of
the bright region, the surface density of stars in bright regions is only
about 4 times that in the field.  Assuming an average O star lives for 10
Myrs and has a velocity relative to the parental molecular cloud of 3 km
s$^{-1}$ (Churchwell 1991), it would travel only $\sim$30 pc over its
lifetime.  Typical dimensions of large field regions are 300$\times$600 pc
(Figure 1b).  Clearly, the field stars cannot all have simply
percolated out of the HII regions we currently see.

Yet another scenario is one where the field stars originated in molecular
clouds located in the field that have since been destroyed so as to leave
the field O stars in a low density gas environment.  The Owens Valley
Millimeter-Wave Interferometer (OVRO) has been used to detect 38 giant
molecular
clouds (masses$\ge$5$\times$10$^4$M$_\odot$) lying within an area of 1.5 to
2.0 kpc$^2$ in M33 (Wilson \& Scoville 1990).  The average surface density of
molecular hydrogen in the OVRO fields is $\Sigma_{H_2}(OVRO)$=7.8M$_\odot$
kpc$^{-2}$ (from Wilson \& Scoville 1990, corrected by a factor of 1.44;
see Thornley \& Wilson 1994) while the average surface density in the inner
kpc of the galaxy is $\Sigma_{H_2}$(1kpc)=6 M$_\odot$ kpc$^{-2}$ (Wilson \&
Scoville 1989, again corrected by a factor of 1.44).  From the numbers of
giant molecular clouds (GMCs) observed, the gas surface density, and the area
covered in the OVRO survey (1.5-2 kpc), we estimate the expected number of
GMCs (with masses$\ge$5$\times$10$^4$M$_\odot$) within the inner kpc of M33
to be of the order of $\sim$50. If the clouds are spread uniformly across a
circular region of radius 1 kpc, the average distance between these clouds
would be 130-150 pc.  Clearly O stars that formed and percolated out from
these large GMCs would not be able to fill the entire area of the field.

It is important to note, however, that the OVRO interferometer detected
only $\sim$40\% of the flux measured in single dish observations of the same
regions of the galaxy (Wilson \& Scoville 1990). The missing flux was
hypothesized to be distributed in a smooth component or dense molecular
clouds less massive than 0.5$\times$10$^5$M$_\odot$ (Wilson \& Scoville
1990).  If we assume that half of the missing flux (30\% of the single dish
flux) is due to small, dense molecular clouds, that the molecular cloud mass
spectrum follows a power law of the form $N(m)\propto m^{-1.5}$ (as
determined from the more massive molecular clouds in M33), and that the minimum
mass of a molecular cloud of interest is 10$^3$ M$_\odot$ (a typical
Taurus-type \lq \lq dark cloud", Goldsmith 1987), then the estimated number
of {\it low-mass} (10$^3$-0.5$\times$10$^5$M$_\odot$) molecular clouds in the
inner kpc of M33 is $\sim$10$^3$. Since the average separation between these
clouds is 30 pc, O stars could easily disperse from these clouds to fill the
entire field.  Therefore, if even half the missing flux is in low-mass
molecular clouds, most of the field O stars could have been born in and
percolated out of (or destroyed) {\it low-mass} molecular clouds.
Observational support for massive star formation in low-mass molecular
clouds can be found within our own Galaxy. A survey of small Galactic HII
regions ($\sim$1/5 the $\half$ luminosity of Orion) and their associated
molecular
clouds (masses$\sim$1-60$\times$10$^3$M$_\odot$) by Hunter \& Massey (1990)
showed that OB star formation has occurred in all of the molecular clouds.
Finally, we point out that these energetic stars may be more effective at
dispersing their lower-mass parental molecular clouds (compared to the more
massive clouds associated with the  bright  HII regions) which could
also account
for the low $\half$ emission in field regions.

 From Table 2  we see that  the relative total $\half$
luminosity per field star is $\sim$1/10$^{th}$ that of a star in a bright
HII region.  This result raises  the question of whether  the field is
\lq \lq leaking" Lyman continuum photons, which presumably escape from the
galaxy. Alternatively, the field population may be older than that in bright
HII regions so that fewer early type stars with large ionizing fluxes
remain. To address these questions, we have determined the stellar
luminosity functions for the three ionized gas environments as well as the
entire sample to see if there is any indication of age segregation (Figure
2).  Weighted least-squares fits to the data to a limiting magnitude of 19
for the bright HII region (to reduce incompleteness effects) and 19.5 for
all others yield slopes of 0.44 $\pm$0.09 (bright), 0.49 $\pm$0.08 (halo),
0.47 $\pm$0.07 (faint) and 0.52 $\pm$0.04 (entire image). The uniformity of
the luminosity function slopes does not suggest that the stellar populations
of the three HII environments have different average ages.  Although
including stars up to V=19.5 in the bright HII regions decreases the slope
to 0.40 $\pm$0.07, the difference is still not significant enough to suggest
a younger population.  As age differences could affect the distribution of
the bright stars most severely, we also examined the distribution of stars
that are brighter than V=18, normalized to the total number brighter than
V=19.  The percentage of stars with V$\leq$18 in the bright, halo and field
regions are respectively, 30 $\pm$10\% , 18 $\pm$7\% and 15 $\pm$5\% .
Clearly again, no region appears to have an over-abundance of luminous stars
and therefore there is no evidence for age segregation between the different
$\half$ emission regions.  It appears therefore that the difference in the
$\half$ luminosity per O star in the field compared to that in the bright
regions is not due to a difference in age but is perhaps indicative that
ionizing radiation from field stars may be escaping out of the surveyed
region.

\section[Ionization Balance in the Inner 1 kpc]
{IONIZATION BALANCE IN THE INNER 1 KPC} \setcounter{equation}{0}

  Star formation rates calculated from $\half$ luminosities assume that all
the Lyman continuum flux emitted by the stars remains in the galaxy (or
region) so that the galaxy (or region) is ionization bounded (eg. Kennicutt
1983).  To test this hypothesis, we have estimated the Lyman continuum and
hence  $\half$ flux emitted by the stars and compared it to that observed in
the $\half$ image.  In carrying out these ionization balance calculations,
we have relied heavily on the models of Panagia (1973) which are applicable
to zero age main sequence stars (ZAMS) as well as main sequence (class V),
luminosity class III and supergiant stars. As an estimate of the
uncertainties in the theoretical models,  we
have also used the blackbody, Kurucz, and  A\&M models.

In all applications of the ionizing radiation models, we have imposed an
upper and lower limit to $\teff$, respectively $T_{eff, U}$ and $T_{eff,
L}$. For comparative purposes, we have set $T_{eff, U}=$60,000 K and
$T_{eff, L}$=30,000 K. The lower limit $T_{eff, L}$ is chosen as the minimum
effective temperature for which all ionization models are defined.  Although
all four models are defined up to $\teff$=50,000 K (for an O4, class V star),
we have chosen $T_{eff,U}=$60,000 K so as to provide a small margin of error
at the high $\teff$ end as well as to roughly account for the Lyman
continuum flux for stars earlier than O4. Stars that had calculated effective
temperatures between 50,000 K and 60,000 K were all assigned Lyman fluxes
corresponding to 50,000 K from the models.  By doing this, we are most
likely under estimating the true Lyman continuum flux from those stars
earlier than O4.  Stars with effective temperatures outside the limits set
by $T_{eff,U}$ and $T_{eff,L}$ have no ionizing flux assigned to them.  We
have not attempted to account for stars with $\teff$$>$$T_{eff,U}$
as these effective temperatures may most likely be the result of photometric
errors (for example, Massey et al. (1989a, 1989b) have found many stars with
unrealistic (B-V) or (U-B) colours the cause of which they attributed to
photometric errors). Similarly, stars with $\teff$$<$$T_{eff,L}$ are also not
accounted for as these stars are most likely to be late type stars (with
correspondingly small ionizing fluxes) or more evolved stars.
We note that the presence of Wolf-Rayet stars is also not taken into
account. While the ionizing Lyman continuum flux from these stars is
estimated to be significant, the exact values are highly uncertain (Massey
et al. 1989b).  Therefore  not accounting for the presence of Wolf-Rayet
stars in the surveyed regions of M33 serves to  underestimate the
predicted theoretical flux. Finally, studies in the LMC and SMC have shown
that effective temperatures calculated from photometry are in general lower
than those determined from spectroscopy (Massey et al. 1989b; Parker et al.
1992) so that here again, predicted ionizing fluxes from photometry are most
likely to be  underestimated.

In terms of the $\half$ fluxes (assuming Case B recombination), the total
flux in $\half$ from all stars with $T_{eff,L}$$\leq$$\teff$$\leq$$T_{eff, U}$
calculated from the four ionization models are (in units of 10$^{39}$ erg
s$^{-1}$ ), F$_{H\alpha}$= 14.46 (Panagia), 15.15 (blackbody), 11.05
(Kurucz) and 13.49 (A\&M). The agreement amongst the different models is
quite good:  the lowest (Kurucz) and highest (blackbody) estimates of
F$_{H\alpha}$  differ by a factor of 1.4.  At low $\teff$, the blackbody
model predicts more Lyman continuum photons than the A\&M and Kurucz models
so that it is not surprising that the total flux from all stars (many of
which have low $\teff$) for the blackbody model is the highest estimate. In
the remainder of the section we will discuss only the $\half$ flux estimated
using the Panagia models.  As the minimum effective temperature for which
these models are defined depends on the luminosity class of the star, i.e.
$\teff$=15,100 K (for class I), $\teff$=16,900 K (for class V and ZAMS) and
$\teff$=16,000 K (for class III), we have defined for each luminosity class
the corresponding $T_{eff, L}$ inferred from the models. As such, the
percentage of stars with (B-V)$\le$0.4 that have $\teff$ between $T_{eff, L}
$ and $T_{eff, U}$ is then 54\% (V$<$18), 60\% (18$<$V$<$19), 74\%
(19$<$V$<$20) and 75\% (20$<$V$<$21).

To obtain the most complete and accurate estimate of the Lyman continuum,
and hence the $\half$, luminosity, we have applied two corrections to the
data. The first concerns incompleteness of the photometric survey at the
faint end of the luminosity function (V$\leq$21). Wilson (1991) determined
that the data were incomplete by 30\% for 19.5$<$V$\leq$20, 55\% for
20$<$V$\leq$20.5 and 65\% for 21$<$V$\leq$21 by comparing the observed
luminosity function with a power law with slope of 0.65. Thus the corrected
fluxes are obtained by scaling the uncorrected fluxes by 1.4 for stars with
19.5$<$V$\leq$20, 2.2 for stars with 20$<$V$\leq$20.5 and 2.9 for stars with
20.5$<$V$\leq$21.  Secondly, we must account for flux from stars that are
below our faint magnitude limit.  We have calibrated a relation between main
sequence mass and M$_V$ using theoretical evolutionary tracks between the
ZAMS and the terminal age main sequence (TAMS) for metallicity Z=0.020 from
Schaller et al. (1992). The tracks were interpolated in steps of 1 M$_\odot$
and then transformed to the observational plane using the equations outlined
in section three of Massey et al. (1989a). As was done in Wilson (1992), we
determined the total range of stellar masses possible for each range in
M$_V$ and assigned to each magnitude bin the average of the maximum and the
minimum possible stellar mass.  At the distance and reddening of M33, V=21
corresponds to 24 M$_\odot$.  Linearly interpolating the spectral type-mass
calibrations given by Popper (1980) and Schmidt-Kaler (1982), we find that
24 M$_\odot$ corresponds to roughly an O7.5-O8 star. We determined the total
number of stars for each spectral (sub)class from O7.5 to B3 (the latest
class for which the Panagia models give theoretical fluxes) using our
spectral type-main sequence mass calibration and the Salpeter initial mass
function (IMF) {\it N(m)d(M)=A$m^{\alpha}dm$} (Salpeter 1955), where
$\alpha$=2.35.  The normalization constant $A$ is determined by integrating
the initial mass function between 24 M$_\odot$ and 65.5 M$_\odot$ and
equating this to the total number of stars for which we could determine
$\teff$, either uncorrected (N$_*\sim$240 (bright), $\sim$400 (halo) and
$\sim$640 (field)) for the conservative flux estimate, or corrected for
incompleteness brighter than V=21 (N$_*\sim$540 (bright), $\sim$1020 (halo)
and $\sim$1280 (field)) for the best flux estimate. We determined the total
flux in each spectral class (or subclass) bin between O7.5 and B3 due to
main sequence stars fainter than V=21 by multiply the average theoretical
flux for that bin (class V assumed) by the total predicted number of stars
in the bin.

  In Table 3 we summarize the observed and predicted $\half$ fluxes, with
theoretical values given in several steps correcting in turn for the effects
discussed above.  Included are the theoretical fluxes predicted from all
stars with $T_{eff, L}$$\leq$$\teff$$\leq$$T_{eff, U}$ (1) for the stars
actually observed (\lq\lq minimum flux"); (2) corrected for stars fainter
than V=21 (\lq\lq conservative flux"); (3) corrected for incompleteness for
both stars brighter and fainter than V=21.  Based on the limited reliability
of the Panagia model in assigning ionizing fluxes from photometric data
alone (Patel \& Wilson 1995), we will consider any agreement between the
observed and predicted $\half$ fluxes that are within a factor of two to be
acceptable.

 If we make no corrections to the data, we find that the predicted fluxes
are in very good agreement with what is observed: over the total image, the
two differ by  less than a factor of 2. The best agreement between theory
and observation is in the bright region ($\sim$1.3:1 correspondence)
followed by the halo region, where the predicted flux is $\sim$1.6 times the
observed flux. The field may be losing some ionizing photons as the
predicted flux is a factor of $\sim$2.4 (Case A) and $\sim$5.7 (Case B)
greater than that observed.

   Including the flux from stars below the faintness limit of the survey
increases the predicted fluxes by a factor of $\sim$2 over both the
individual ionized gas regions as well as the entire survey area. The total
theoretical flux exceeds that observed by factors of 2.1 (bright), 3.6
(halo), 5.4 (field: Case A), 12.9 (field:  Case B), 3.1 (total image: Case
A) and 4.1 (total image: Case B).  Clearly adopting Case B recombination for
the field results in predicted ionizing fluxes that are substantially larger
than observed. Including stars below the faintness limit of our survey is
the minimum correction we must make to our data and thus these fluxes
represent a lower limit to the theoretically predicted $\half$ fluxes.

 Correcting the data for incompleteness (V$\leq$21) increases the predicted
fluxes by a factor of less than 2 over the different ionized gas
environments.  The most complete estimate of the predicted flux includes
both incompleteness corrections for V$\leq$21 and main sequence stars with
V$>$21.  From Table 3 we see that over the entire surveyed
region, these final fluxes are a factor of $\sim$6.7 (Case A) to $\sim$9.3
(Case B) greater than observed.  The best agreement is again found in the
bright regions where the predicted flux is $\sim$4 times that observed and
the worst agreement is in the field where the predicted flux exceeds the
observed by factors of $\sim$13 (Case A) to $\sim$31 (Case B).  This most
generous estimate of the theoretical fluxes after corrections and the most
conservative estimate differ by less than a factor of 3 from  each other.

The best agreement between observations and theory in the field, as well as
for the entire surveyed region, is obtained when we assume Case A (optically
thin) recombination.  Bright HII regions are likely to have larger opacities
than faint regions where gas densities are likely to be smaller. Although we
have no direct measure of the density of gas in the various regions, we can
place limits on this quantity by comparing the observed $\half$ surface
brightness with that expected for HII regions of various different gas
densities.  If we consider a B0 star ($L_{H\alpha}$=5.82$\times$10$^{35}$
erg sec$^{-1}$ and $N_{LyC}$=4.27$\times$10$^{47}$ photons sec$^{-1}$;
Panagia 1973) located in the field, then the maximum surface brightness
observed in the field (I$_F$=3.61$\times$10$^{-16}$ erg s$^{-1}$cm$^{-2}$
arcsec$^{-2}$) limits the gas density to be $\le$5 cm$^{-3}$. We have used
the Stromgren radius (eg. Spitzer 1978) to estimate the radius of the HII
region and assumed the gas within the Stromgren sphere of the star is
completely ionized.  For the HII region of a field O5 star to have a surface
brightness equal to I$_F$, the density of gas would have to be $\le$0.5
cm$^{-3}$.  In contrast, if the surface brightnesses of HII regions formed
by B0 and O5 stars are to equal the average surface brightness observed in
bright HII regions (I$_B$=3.44$\times$10$^{-15}$ erg s$^{-1}$cm$^{-2}$
arcsec$^{-2}$), the gas density must be $\sim$15 cm$^{-3}$ (B0) and $\sim$1
cm$^{-3}$ (O5).  Thus the density of gas in the field is likely to be lower
than that in the bright HII regions which lends some support to our
assumption of Case A recombination in the field and Case B elsewhere in the
galaxy.

Comparing the corrected final theoretical fluxes with the observed fluxes
{\it when the uniform background has not been removed} reveals that the two
still differ by up to a factor $\sim$2-5. Clearly the field region is most
sensitive to the inclusion or exclusion of the uniform background: the field
could be in ionization balance if Case A recombination is assumed and the
uniform background is not removed.  However, including the background and
assuming Case A recombination where appropriate, comparison of the
theoretical best estimate fluxes with the observed fluxes suggests that {\it
over the total, as well as the bright and halo regions surveyed, the galaxy
is not in a state of ionization balance.}

 With so much ionizing flux missing we are left with the question:  where
does it go? One possible fate for leaked photons is in powering the halo
ionized gas.  The low density and expected small opacities in the field
combined with the enormous reservoir of leaked flux provided by the OB stars
would clearly increase the chances for the photons to reach the large
distances above the plane before recombining and producing the halo
ionized gas.  Additionally, the large pool of ionizing photons available
would be advantageous in producing a halo   ionized gas with a total
luminosity that is a sizable fraction of the total $\half$ emission measured
in the galaxy.

 Star formation rates from $\half$ luminosities are based on the assumption
that all the ionizing radiation from stars remains in the galaxy. If a
significant portion of the ionizing flux is leaking out of the galaxy, SFRs
calculated from observed $\half$ fluxes would underestimate the true SFR.
Our results suggest that this is the case in M33. We find, in fact, that SFRs
calculated from observed $\half$ luminosities alone underestimate those
calculated from the most realistic predicted fluxes (best estimate scenario
and assuming Case A recombination in the field) by a factor of $\sim$4
(bright), $\sim$8 (halo), $\sim$13 (field) and $\sim$7 (total image). The
implications of this result for SFR calculations in other systems could be
serious. Clearly much will depend on how ionizing photon leakage correlates
with other characteristics of a galaxy such as its morphology, age and the
state of the interstellar medium. If M33 is typical of its class, then
applying this technique in late-type spiral galaxies could severely
underestimate the actual SFR.  The outcome for irregular galaxies (due
to their patchy gas distribution) and elliptical galaxies (due to their low
gas densities) may be comparable to that found here for M33 but
would certainly need to be verified by a similar type of investigation
before any conclusions can be drawn.

\subsection [Ionization Balance in an Isolated Region]
{IONIZATION BALANCE IN AN ISOLATED REGION}

 Having looked at the question of ionization balance on the large scale, we
consider now the small scale and focus on one isolated region of the galaxy
containing both bright and halo HII emission. The chosen region, denoted
Region A, lies in the south-east portion of the galaxy, and covers an area of
$\sim$300$\times$500 pc (90$^{\prime\prime}\times$130$^{\prime\prime}$;
Figure 3). We have photometry for 19 stars located in the region
with bright $\half$ emission and 17 stars in the halo emission region.
Table 4 summarizes the observed and theoretical $\half$ fluxes
in the bright, halo and combined (bright and halo) regions.  We find that
observations and model predictions are best matched for the simplest
situation where no corrections are made: the predicted fluxes in the bright
and halo regions are, respectively, a factor 1.8 and 0.1 times that observed
so that the region as a whole is in balance.  Correcting for stars fainter
than V=21 increases the predicted fluxes by factors of $\sim$2 in the
bright, $\sim$14 in the halo region and $\sim$1.4 over the combined region.
This result can be considered to be the most conservative corrected estimate
of the theoretical flux in Region A.  Correcting for incompleteness (V$\le$21)
increases the minimum theoretical flux by a factor of $\sim$3 over both the
bright and halo regions.  The best estimate of the predicted fluxes are a
factor of 8 (bright), 5 (halo) and 7 (combined bright+halo) greater than that
observed.  Therefore, only in the simplest situation of the uncorrected data
is Region A close to being ionization bounded.  In the more realistic case
where the data has been corrected for incompleteness in stars brighter than
V=21 and flux from stars fainter than V=21, this region is {\it clearly not
ionization bounded.} Even the conservative estimate of the corrected flux
indicates this region is leaking photons to its surroundings.  Furthermore,
including the uniform background cannot bring the combined bright and halo
emission regions into ionization balance.  Thus our results for a small,
individual region confirm those obtained for the larger surveyed area:
ionizing flux from OB stars is found to be missing on both large and small
scales in M33.

\section
[Conclusions] {CONCLUSIONS}

Using $\half$ data and UBV photometry of blue stars, we have investigated
the distribution of HII regions and OB stars and tested the hypothesis of
ionization balance within the inner kpc of M33.  The main results are
summarized below.

(1) The $\half$ emission appears to be distributed in three distinct
components, denoted bright, halo  and field, that are spread out over the
galaxy. Roughly one third of the brightest blue stars (V$\leq$18) are found
in the field region.  Including all stars within the photometric limit V=21,
we find that $\sim$50\% of the OB stars (luminous blue stars with
$\bv$$\leq$0.4) are located in the field. From this result we estimate
50\% of the main sequence lifetime of an OB star is spent outside a
recognizable HII region. The implication of this for molecular cloud
lifetimes is that if the field stars escape from bright HII regions by
destroying their parent molecular clouds, then molecular cloud lifetimes
must be shorter than $\sim$10$^7$ yrs. On the other hand, if OB stars simply
move out of the clouds, molecular cloud lifetimes could be considerably
longer.

(2) Assuming typical OB star velocities are $\sim$3 km $s^{-1}$ we find
field stars could not all have originated in and percolated out of existing
HII regions or for that matter, a strictly massive
($\ge$0.5$\times$10$^5$M$_\odot$) GMC population. If the
galaxy contains GMCs with masses down to $\sim$10$^3$M$_\odot$ (typical
Taurus-type dark clouds), there would exist sufficient numbers of GMCs to
make the average separation between them comparable to the maximum distance
an OB star can travel over its lifetime.  Therefore, the field star
population could have originated from and percolated out of a GMC population
with masses $\ge$10$^3$M$_\odot$.

(3) The slopes of the luminosity functions for stars in the bright, halo and
field regions are comparable to one another, which suggests that the stellar
populations in the different ionized gas environments must have similar ages.
This result is supported by the brightest (V$\leq$18) blue stars which show
no evidence of being concentrated in any one type of ionized gas environment.

(4) We have estimated the theoretical ionizing flux from the observed OB
stars using the Panagia (1973) models and converted these to theoretical
$\half$ fluxes assuming Case B recombination for the bright and halo regions
and Case A and B recombination for the field region.  Ionization balance
calculations show that the inner region of M33 {\it is not in ionization
balance}:  the observed fluxes are a factor of $\sim$7-9 (Case A-Case B)
less than the predicted best estimate fluxes and a factor of $\sim$3-4 (Case
A-Case B) less than the predicted conservative estimate fluxes. The greatest
loss of ionizing flux occurs in the field region:  even in the conservative
case, the predicted $\half$ flux is a factor of 5 (Case A recombination)
greater than that observed. The discrepancy between the two increases to a
factor of 13 (Case A recombination) for the best estimate case of the
predicted flux. We find  Case A recombination provides the best
agreement between observed and predicted $\half$ fluxes in the field. Of the
three ionized gas environments, the bright regions are closest to being
ionization bounded.  Investigation of an individual, isolated region
containing both bright and halo emission produces comparable results:  the
region is unbalanced by a factor of $\sim$3 and $\sim$7 for respectively, the
conservative and best estimate cases.

(5) Based on our result that the separate ionized gas environments and
individual regions of the galaxy are leaking ionizing flux, SFRs calculated
from $\half$ luminosity must underestimate the true SFR.  Over the entire
surveyed region, the difference between the true SFR and that calculated
from $\half$ could differ by a factor of $\sim$3-17.

\clearpage

\center{\bf FIGURE LEGENDS}

\begin{description}

\item[Figure 1a.]

The continuum subtracted $\half$ image of the central
8.7$^{\prime}$$\times$8.7$^{\prime}$ of M33 illustrating the distribution of
the bright, halo and field $\half$ emission regions.  North is at the top
and east is to the left.  The contours, corresponding to surface
brightnesses of {\it I}=1.74$\times$10$^{-15}$ and 4.57$\times$10$^{-16}$ erg
s$^{-1}$cm$^{-2}$arcsec$^{-2}$, separate,  respectively, the bright regions
from the halo and the halo regions from the field. For orientation, the
giant HII region NGC 595 is located on the top right-hand side of the image
and the center of the galaxy coincides roughly with that of the image.

\item[Figure 1b.]

The distribution of all
field OB stars (V$\leq$21 and (B-V)$\leq$0.4) in the inner kpc of M33.  As
photometric coverage of the two vertical strips at the far left and right
edges of the image was not available, these regions were excluded from the
analysis.

\item[Figure 2.]

The luminosity functions
for stars in the entire image as well as the bright, halo and field HII
regions.  An arbitrary offset in the ordinate has been applied for the
purposes of clarity.  Overlaid on the data are the weighted least squares
fits (to a limiting magnitude of V=19 for the bright region and V=19.5
elsewhere).

\item[Figure 3.]
Expanded View of Region A which was selected for individual investigation.
Overlaid are the stars in the bright and halo $\half$ emission areas within
the region.

\end{description}
\clearpage

\begin{table*}
\begin{center}
\begin{tabular}{lcccccccc}
\tableline
\tableline
\multicolumn{1}{c}{
V MAG   \tablenotemark{*}}
&  \multispan2\hfil  BRIGHT  \hfil
&  \multispan2\hfil  DIFFUSE  \hfil
&  \multispan2\hfil FIELD   \hfil
&\multispan2\hfil   ENTIRE IMAGE\hfil\\
&   \# of stars & \% in  region &   \# of stars & \% in  region
&   \# of stars & \% in  region & \# of stars & \% in  region \\
\tableline
\tableline
V  $<$18 &
 22 $\pm$5 & 5.6  $\pm$1.2\%  &  15  $\pm$4  & 2.2 $\pm$0.6\% &
 18 $\pm$4 & 1.6  $\pm$0.4\% &  55  $\pm$7  & 2.5  $\pm$0.3\% \\
18 $<$V$<$19   & 50  $\pm$7  & 13 $\pm$2\%  & 66  $\pm$8  & 10 $\pm$1\%
               & 100 $\pm$10 &  9 $\pm$1\%  & 216 $\pm$15 & 10 $\pm$1\% \\
19$<$V$<$20    & 123 $\pm$11 & 32  $\pm$3\% & 173 $\pm$13 & 25  $\pm$2\%
               & 276 $\pm$17 & 25 $\pm$1\%  & 572 $\pm$24 & 26 $\pm$ 1 \%\\
20$<$V$<$21    & 195 $\pm$14 & 50  $\pm$4\% & 434 $\pm$21 & 63   $\pm$4\%
               & 708 $\pm$27 & 64  $\pm$3\% & 1337$\pm$37 & 61 $\pm$2\% \\
\tableline
V$\leq$21(Total)&390 $\pm$20 &  &  688 $\pm$26 & & 1102 $\pm$33 & & 2180 \\
\tableline
\tableline
\end{tabular}
\end{center}

\tablenotetext{*}{Note : Only stars with (B-V)$\leq$0.4}
\caption{The Distribution of Blue  Stars in Three Ionized Gas Environments
in M33.}
\label{tbl-numregion}
\end{table*}

\clearpage

\begin{table*}
\begin{center}
\begin{tabular}{lcccc}
\tableline
\tableline
 &  BRIGHT & HALO  &FIELD &ENTIRE IMAGE \\
\tableline
\tableline
Total \# of Stars M$_V$$\leq$21 & 390& 688&1102&2180\\
Area Covered  (kpc$^2$)  & 0.22 & 0.83 &2.28 & 3.33 \\
Surface Density of Stars(kpc$^{-2}$)  & 1770 & 829 & 483  & 655 \\
\multicolumn{1}{l}{Average Surface Brightness [I] \tablenotemark{a}} & 3.44&
0.62& 0.09& 4.15 \\
\multicolumn{1}{l}
{Total Luminosity [L$_{ob}$]  \tablenotemark{b}}   &3.78 & 2.57 &1.01 & 7.36 \\
\multicolumn{1}{l}{Relative  Luminosity per Star \tablenotemark{c}}
  & 1.0 & 0.38 & 0.10  & 0.35 \\
\tableline
\tableline
\end{tabular}
\end{center}




\caption{Average Properties of the Three Ionized Gas Environments.}
\label{tab-ch2-2}
\end{table*}

$^{a}$ In units of 10$^{-15}$ erg s$^{-1}$cm$^{-2}$
arcsec$^{-2}$.This is measured above the average background.

$^{b}$  In units of L$_{39}$$\equiv$10$^{39}$ erg s$^{-1}$.
L$_{ob}$ is measured above the average background luminosity L$_{bg}$= 4.0
$\pm$ 0.08 $\times$10$^{34}$ erg s$^{-1}$arcsec $^{-2}$. The total observed
luminosity L$_{tot}$= L$_{ob}$+L$_{bg}$, for the bright, halo, field and
entire image are, respectively, 3.9$\times$10$^{39}$, 4.6$\times$10$^{39}$,
7.0$\times$10$^{39}$ and 15.6$\times$10$^{39}$ erg s$^{-1}$.

$^c$ In units of 0.97$\times$10$^{37} $ erg s$^{-1}$ star$^{-1}$.
In the case where the average background was not subtracted off, the
corresponding numbers are: 1., 0.67, 0.64 and 0.72 (in units of
1.0$\times$10$^{37}$ erg s$^{-1}$) for, respectively, the bright, halo,
field and collective  regions.

\clearpage

\begin{table*}
\begin{center}
\begin{tabular}{lcccc}
\tableline
\tableline
                &  BRIGHT       & HALO  & FIELD         & TOTAL \\
                &               &               & (Case A/ B)   &(Case A/ B)\\
\tableline
\multicolumn{1}{l}{Minimum Flux\tablenotemark{a}}
 & 4.90 & 4.07  & 2.40 / 5.75   & 11.48 / 14.79 \\
\multicolumn{1}{l}{Incompleteness (V$\le$21)\tablenotemark{b}}
& 8.13 & 8.32 &       4.75 /  11.48 & 21.20 / 27.93 \\
\tableline
\multicolumn{1}{l}{Best Estimate \tablenotemark{c}}
 & 15.03 & 21.47 & 13.23/31.88 & 49.73 / 68.38 \\
\multicolumn{1}{l}{Conservative  Estimate \tablenotemark{d}}
& 7.92 & 9.32 &   5.42 / 13.00 & 22.66 / 30.24 \\
\tableline
\multicolumn{1}{l}{Observed Flux\tablenotemark{e}}
& 3.78  & 2.57          &1.01   & 7.36\\
\tableline
\tableline

\end{tabular}
\end{center}

\caption {Observed and Predicted  H$_\alpha$ Flux
in  the Three   Ionized Gas  Environments.}
\label{tab-ch2-3}
\end{table*}

$^a$ Minimum: no corrections applied.
All fluxes in  units of L$_{39}\equiv$10$^{39}$ erg s$^{-1}$
arcsec$^{-2}$.

$^b$ Incompleteness:  corrected
for  incompleteness for  stars brighter than V=21.

$^c$ Best estimate:  corrected for
incompleteness  at  bright (V$\leq$21) and faint (V$>$21) magnitudes.

$^d$ Conservative estimate: corrected
only for stars with  V$>$21.

$^e$ Observed fluxes do not include the average sky
background. If the average background is included,  the relevant
numbers are: 3.9 (bright), 4.6 (halo), 7.0 (field) and 15.6 (entire
image).

\clearpage

\begin{table*}
\begin{center}
\begin{tabular}{lccc}
\tableline
\tableline
                &  BRIGHT       & HALO  & COMBINED \\
                &               &               & (BRIGHT $+$ HALO)\\
\tableline
\multicolumn{1}{l}{Minimum Flux\tablenotemark{a}}
 & 0.24         & 0.01  & 0.25 \\
\multicolumn{1}{l}{Incompleteness (V$\leq$21)\tablenotemark{a}}
& 0.69  &0.03           & 0.72 \\
\tableline
\multicolumn{1}{l}{Best Estimate \tablenotemark{a}}
 & 1.00 & 0.37 & 1.37 \\
\multicolumn{1}{l}{Conservative  Estimate \tablenotemark{a}}
& 0.40 & 0.14 & 0.54 \\
\tableline
\multicolumn{1}{l}{Observed Luminosity [L$_{ob}$]\tablenotemark{b}}
& 0.13  &0.07           &0.20 \\
\# Stars Observed [$S_{ob}$]  & 19              & 17            &36\\
\# Stars with $T_{eff, L}<\!t_{eff}<T_{eff, U}$
&       11      &14     &25     \\
\multicolumn{1}{l}{ L$_{ob}$/$S_{ob}$\tablenotemark{c}}
 & 1 &  0.64& 0.82 \\
\tableline
\tableline

\end{tabular}
\end{center}

\caption {The Observed and Predicted H$_\alpha$ Fluxes
and OB Star Content in Region A.}

\label{tab-ch2-4}
\end{table*}

$^a$ Predicted $\half$ luminosity accounting for incompleteness
(as in Table 3).  All fluxes in units of L$_{39}\equiv$10$^{39}$ erg s$^{-1}$
arcsec$^{-2}$.

$^b$  Observed luminosities have been  background subtracted.
If the background is included, the numbers are:  0.18
(bright) and 0.14 (halo) and 0.32 (combined bright and halo).

$^c$ In units of 6.8$\times$10$^{36}$erg s$^{-1}$ arcsec$^{-2}$
star$^{-1}$.

\clearpage

\end{document}